# Intrinsic localized modes in a two-dimensional checkerboard ferromagnetic lattice

Wenhui Feng[a], Heng Zhu[a], Bing Tang[*]

*Department of Physics, Jishou University, Jishou 416000, China*

ABSTRACT

An analytical work on intrinsic localized modes in a two-dimensional Heisenberg ferromagnet on the checkerboard lattice is presented. Taking advantage of an asymptotic method, the governing lattice dynamical equations are reduced to one (2+1) -dimensional nonlinear Schrödinger. In our work, we obtain two types of nonlinear localized mode solutions, namely, Brillouin zone center modes and Brillouin zone corner modes. The occurrence conditions for these intrinsic localized modes are given in detail. Especially, we find that the competition between the Dzialozinskii-Moriy interaction and the next-nearest neighbor interaction of the checkerboard ferromagnet has an effect on the local structure of the Brillouin zone corner acoustic mode.

## 1. Introduction

Intrinsic localized modes ( also referred to as discrete breathers ) were for the first time found by Taken and Sievers in a 1D β-Fermi–Pasta–Ulam lattice chain[1]. Soon afterwards, numerous works on the intrinsic localized mode in 1D discrete nonlinear lattice systems have been reported[2]. Especially, the existences of the intrinsic localized mode in those one dimensional nonlinear lattice systems have been rigorously proven via using an anti-continuous limit[3-5]. It has been shown that both the nonlinearity and the discreteness are of vital importance to the emergence of such intrinsic localized modes. Experimentally, the intrinsic localized mode has been detected in the antiferromagnetic material[6], the Josephson array[7], and the diatomic electrical lattice[8].

[*]Bing Tang:bingtangphy@jsu.edu.cn
These authors contributed equally to this work.

Over the past decades, more and more researchers also have paid attention to the intrinsic localized mode in those higher dimensional lattices[9-20]. In the early paper, Takeno putted forward lattice Green's functions to seek for localized mode solutions in 1D, 2D and 3D nonlinear lattices[9]. By use of a similar method, he has studied the characteristic and profile of the localized mode in general $d$-dimensional nonlinear lattices[10]. Afterwards, Wattis *et al.* have obtained analytical forms of the intrinsic localized mode in some two-dimensional Fermi–Pasta–Ulam lattices via using an asymptotic method based on the small amplitude ansatz[11-13]. In their works, two-dimensional square, honeycomb, and triangular lattice structures have been respectively considered. On the other hand, some authors have used numerical methods to obtain intrinsic localized mode solutions in various two-dimensional nonlinear lattice systems[14-20]. For example, Marin *et al.* have performed extensive numerical works on intrinsic localized modes in two-dimensional nonlinear lattices[15-17], and English *et al.* have showed that numerical simulations is in good agreement with experimental results for the intrinsic localized mode in 2D electrical square and honeycomb electrical lattices[18]. Recently, Watanabe and Izumi have obtained numerical solutions of intrinsic localized modes in a 2D hexagonal Fermi–Pasta–Ulam nonlinear lattice system by using a generalized minimal residual method[19]. Meanwhile, Babicheva *et al.* have proposed a numerical scheme based on the localizing function to the search for the asymptotic intrinsic localized mode solution in one 2D triangular β-Fermi-Pasta-Ulam-Tsingou nonlinear lattice system[20]. Furthermore, Dai *et al.* have investigated numerically the emergence of the intrinsic localized mode and the corresponding energy localization in a two-dimensional nonlinear honeycomb lattice derived from the grapheme [21].

In this article, we analytically investigate existences and properties of intrinsic localized modes in a two-dimensional checkerboard Heisenberg ferromagnetic spin lattice by using an asymptotic method developed by Butt and Wattis[13]. In our previous works, we have applied this analytic scheme to obtain some intrinsic localized mode solutions in the two-dimensional square, hexangular, and honeycomb Heisenberg ferromagnetic lattice systems[22-24]. Theoretically, the checkerboard lattice structure can be regarded as the 2D counterpart of one 3D pyrochlore lattice structure [25]. In the linear spin wave approximation, topological properties of magnons have been well studied in the two-dimensional checkerboard Heisenberg ferromagnet[26,27]. The goal of the present research is to explore that the interaction between magnons may cause a nonlinear localized excitation, i.e., the intrinsic localized mode. More particulars will be exhibited in the following sections.

**2. The lattice model Hamiltonian and its quantization**

In the present research, we take into account an anisotropy Heisenberg ferromagnet placed on the 2D checkerboard lattice. Under an external magnetic field, the spin Hamiltonian of this 2D checkerboard ferromagnet reads

$$H = -J_1 \sum_{\langle i,j \rangle} \vec{S}_i \cdot \vec{S}_j - J_2 \sum_{\langle\langle i,j \rangle\rangle} \vec{S}_i \cdot \vec{S}_j + D \sum_{\langle i,j \rangle} v_{ij} \left( S_i^x S_j^y - S_i^y S_j^x \right)$$
$$- A \sum_i \left( S_i^z \right)^2 - g\mu_B H_{\text{ext},z} \sum_i S_i^z.$$

(1)

We signify the first-nearest neighbor spins with interactions $J_1$ and $D$ between spins at lattice sites A and B via $\langle i,j \rangle$. What is more, $\langle\langle i,j \rangle\rangle$ stands for second-nearest neighbor exchange interactions between lattice sites A and A, and B and B, with $J_2$ along those solid lines. The third term corresponds to the first-nearest neighbor (Dzyaloshinskii-Moriya) DM interaction, where $\vec{D}_{ij}$ is referred to as the DM interaction vector between nearest neighbor lattice sites $i$ and $j$. According to the Moriya rule, one can write down $\vec{D}_{ij} = v_{ij} D \vec{e}_z$, where $v_{ij}$ is in fact an orientation-dependent coefficient[26]. The last term represents the Zeeman coupling with one applied magnetic field $\mathbf{H} = H_{\text{ext},z} \mathbf{e_z}$. As can be clearly seen in Fig. 1, this 2D checkerboard lattice possesses two inequivalent lattice sites A and B, which lie on a subset of one underlying square lattice. For the sake of depicting the position of the $(m,n)$ lattice site for this square lattice, one can use a pair orthonormal basis vectors $\mathbf{e_x}$ and $\mathbf{e_y}$ so that those position vectors can be represented as $m\mathbf{e_x} + n\mathbf{e_y}$. In order to specify this two-dimensional checkerboard ferromagnetic lattice, we need to reserve only those lattice sites $(m,n)$, which meet the conditions $m + n = even$. In Fig. 1, those color solid circles stand for the sites reserved in this two-dimensional checkerboard lattice that satisfy these conditions, and those black open circles hold up all the rest of lattice sites in the square lattice.

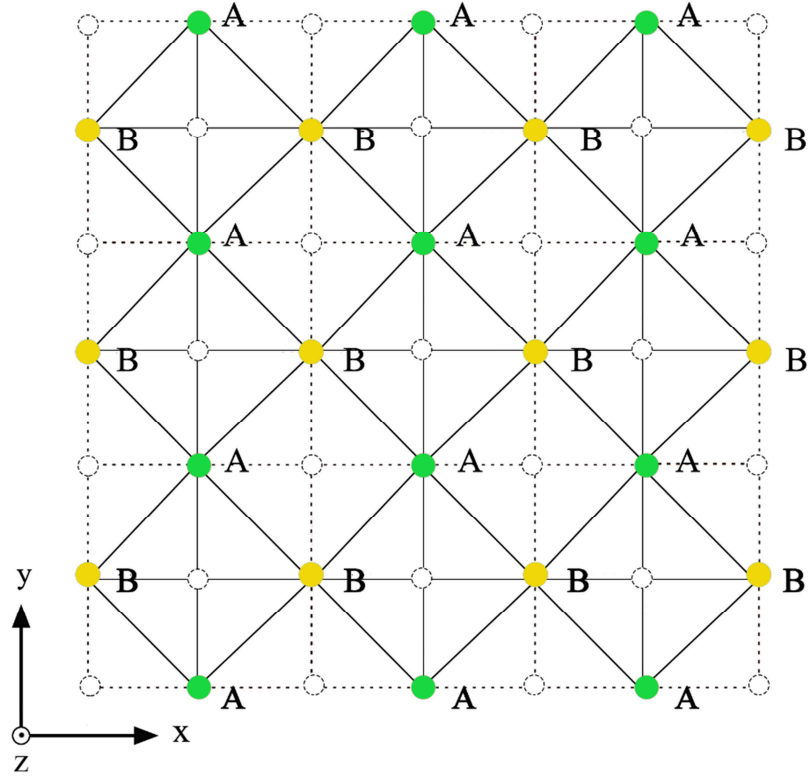

**Fig. 1.** (Color online) Diagrammatic drawing of the checkerboard lattice arranged on the x–y plane. Color solid circles are those lattice sites in the 2D checkerboard lattice, black open circles correspond to the unused lattice sites in the fundamental rectangular lattice. The dotted lines stand for those unit cells, each of which owns two nonequivalent lattice sites A and B.

Here, lattice sites need to be reindexed via considering a rectangular lattice, whose basis vector is $\mathbf{R} = \{\mathbf{e}'_x, \mathbf{e}'_y\}$ with $\mathbf{e}'_x = \mathbf{e}_x = [2,0]^T$ and $\vec{e}'_y = 2\vec{e}_y^{\,T} = [0,2]^T$, as shown in Fig. 2. We only consider half of the $(m,n)$ indices so as to ensure that the sum $m+n$ is even. Let us set an origin with the coordinate $(0,0)$, thus the position vector for the lattice site $(m,n)$ is written as $\mathbf{r}_{m,n} = m\mathbf{e}'_x + n\mathbf{e}'_y$.

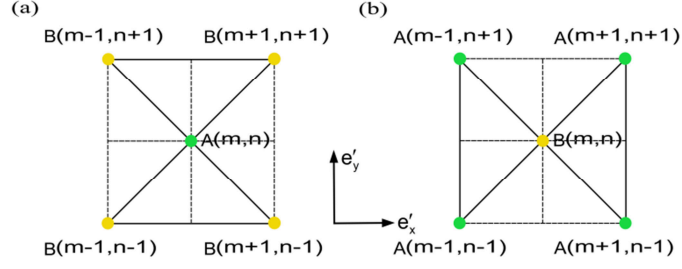

**Fig. 2.** Labelling of sites in the two-dimensional checkerboard lattice with the basis vector $\mathbf{R} = \{\mathbf{e}'_x, \mathbf{e}'_y\}$.

To bosonize the checkerboard Heisenberg ferromagnet Hamiltonian(1), we use the Dyson-Maleev transformations [28,29] for the two sublattices A and B, which are given by

$$S^{A+} = S^{Ax} + iS^{Ay} = \sqrt{2S}(1 - \frac{a^+ a}{4S})a, \tag{2a}$$

$$S^{A-} = S^{Ax} - iS^{Ay} = \sqrt{2S}a^+, \tag{2b}$$

$$S^{Az} = S - a^+ a \tag{2c}$$

(similarly for sublattice B).

When Glauber coherent-statere presentation[30] is adopted, we can suppose that the system state vector $|\Psi(t)\rangle$ has the following form

$$|\Psi(t)\rangle = \left(\prod_{A\text{-sublattice}} |\alpha_{m,n}\rangle\right)\left(\prod_{B\text{-sublattice}} |\beta_{m,n}\rangle\right). \tag{3}$$

By utilizing the variational method, one can derive the equations of motion for coherent-state amplitudes $\alpha_{m,n}$ and $\beta_{m,n}$. Through some simple calculations, we find that the equations of motion for A sites in arrangement 1 is

$$i\frac{\partial \alpha_{m,n}}{\partial t} = \omega_0 \alpha_{m,n} - J_1 S(\beta_{m+1,n+1} + \beta_{m+1,n-1} + \beta_{m-1,n+1} + \beta_{m-1,n-1}) - J_2 S(\alpha_{m,n+2} + \alpha_{m,n-2})$$

$$+ iDS(\beta_{m+1,n+1} - \beta_{m-1,n+1} - \beta_{m+1,n-1} + \beta_{m-1,n-1}) + \frac{J_1}{4}(\beta^*_{m+1,n+1}\alpha^2_{m,n} + |\beta_{m+1,n+1}|^2 \beta_{m+1,n+1}$$

$$- 4\alpha_{m,n}|\beta_{m+1,n+1}|^2 + \beta^*_{m+1,n-1}\alpha^2_{m,n} + |\beta_{m+1,n-1}|^2 \beta_{m+1,n-1} - 4\alpha_{m,n}|\beta_{m+1,n-1}|^2$$

$$+ \beta^*_{m-1,n+1}\alpha^2_{m,n} + |\beta_{m-1,n+1}|^2 \beta_{m-1,n+1} - 4\alpha_{m,n}|\beta_{m-1,n+1}|^2 + \beta^*_{m-1,n-1}\alpha^2_{m,n}$$

$$+ |\beta_{m-1,n-1}|^2 \beta_{m-1,n-1} - 4\alpha_{m,n}|\beta_{m-1,n-1}|^2) + \frac{J_2}{4}(\alpha^*_{m,n+2}\alpha^2_{m,n} + |\alpha_{m,n+2}|^2 \alpha_{m,n+2} - 4\alpha_{m,n}|\alpha_{m,n+2}|^2$$

$$+ \alpha^*_{m,n-2}\alpha^2_{m,n} + |\alpha_{m,n-2}|^2 \alpha_{m,n-2} - 4\alpha_{m,n}|\alpha_{m,n-2}|^2) - \frac{iD}{4}(|\beta_{m+1,n+1}|^2 \beta_{m+1,n+1} - \beta^*_{m+1,n+1}\alpha^2_{m,n}$$

$$- |\beta_{m-1,n+1}|^2 \beta_{m-1,n+1} + \beta^*_{m-1,n+1}\alpha^2_{m,n} - |\beta_{m+1,n-1}|^2 \beta_{m+1,n-1} + \beta^*_{m+1,n-1}\alpha^2_{m,n} + |\beta_{m-1,n-1}|^2 \beta_{m-1,n-1}$$

$$- \beta^*_{m-1,n-1}\alpha^2_{m,n}) - 2A|\alpha_{m,n}|^2 \alpha_{m,n}$$

(4)

with $\omega_0 = (2S-1)A + g\mu_B H_{ext,z} + 4J_1 S + 2J_2 S$. In a similar way, it is easy to obtain the equations of motion for B sites in arrangement 2, namely,

$$i\frac{\partial \beta_{m,n}}{\partial t} = \omega_0 \beta_{m,n} - J_1 S(\alpha_{m+1,n+1} + \alpha_{m+1,n-1} + \alpha_{m-1,n+1} + \alpha_{m-1,n-1}) - J_2 S(\beta_{m+2,n} + \beta_{m-2,n})$$

$$- iDS(\alpha_{m+1,n+1} - \alpha_{m+1,n-1} - \alpha_{m-1,n+1} + \alpha_{m-1,n-1}) + \frac{J_1}{4}(|\alpha_{m+1,n+1}|^2 \alpha_{m+1,n+1} + \alpha^*_{m+1,n+1}\beta^2_{m,n}$$

$$- 4|\alpha_{m+1,n+1}|^2 \beta_{m,n} + |\alpha_{m+1,n-1}|^2 \alpha_{m+1,n-1} + \alpha^*_{m+1,n-1}\beta^2_{m,n} - 4|\alpha_{m+1,n-1}|^2 \beta_{m,n} + |\alpha_{m-1,n+1}|^2 \alpha_{m-1,n+1}$$

$$+ \alpha^*_{m-1,n+1}\beta^2_{m,n} - 4|\alpha_{m-1,n+1}|^2 \beta_{m,n} + |\alpha_{m-1,n-1}|^2 \alpha_{m-1,n-1} + \alpha^*_{m-1,n-1}\beta^2_{m,n} - 4|\alpha_{m-1,n-1}|^2 \beta_{m,n})$$

$$+ \frac{J_2}{4}(\beta^*_{m+2,n}\beta^2_{m,n} + |\beta_{m+2,n}|^2 \beta_{m+2,n} - 4\beta_{m,n}|\beta_{m+2,n}|^2 + \beta^*_{m-2,n}\beta^2_{m,n} + |\beta_{m-2,n}|^2 \beta_{m-2,n} - 4\beta_{m,n}|\beta_{m-2,n}|^2)$$

$$+ \frac{iDS}{4S}(|\alpha_{m+1,n+1}|^2 \alpha_{m+1,n+1} - \alpha^*_{m+1,n+1}\beta^2_{m,n} - |\alpha_{m+1,n-1}|^2 \alpha_{m+1,n-1} + \alpha^*_{m+1,n-1}\beta^2_{m,n}$$

$$- |\alpha_{m-1,n+1}|^2 \alpha_{m-1,n+1} + \alpha^*_{m-1,n+1}\beta^2_{m,n} + |\alpha_{m-1,n-1}|^2 \alpha_{m-1,n-1} - \alpha^*_{m-1,n-1}\beta^2_{m,n}) - 2A|\beta_{m,n}|^2 \beta_{m,n}.$$

(5)

## 4. Asymptotic analysis and nonlinear amplitude equation

Here, we shall make use of an asymptotic method developed by Butt and Wattis to seek the localized solution to the Eqs. (7) and (8). First, we need to rescale the present variables $m$, $n$ and $t$ via considering the following new variables

$$x = \rho m, \quad y = \rho n, \quad \tau = \rho t, \quad T = \rho^2 t, \tag{6}$$

where $\rho \ll 1$ is a small quantity.

In principle, different ansatzes should be respectively applied to the A and B sites. For A sites, we consider that their trial solutions possess the following form

$$\alpha_{m,n}(t) = \rho e^{i\phi} f(X,Y,\tau,T) + \rho^2 [g_0(X,Y,\tau,T) + e^{i\phi} g_1(X,Y,\tau,T) + e^{2i\phi} g_2(X,Y,\tau,T)]$$
$$+ \rho^3 [h_0(X,Y,\tau,T) + e^{i\phi} h_1(X,Y,\tau,T) + e^{2i\phi} h_2(X,Y,\tau,T) + e^{3i\phi} h_3(X,Y,\tau,T)] + \ldots$$

(7)

Here, $\phi = k_x m + k_y n - \omega t$ signifies the phase of the carrier wave. Note that the wavevector **k** has been written as $\mathbf{k} = k_x \mathbf{e}_x + k_x \mathbf{e}_y$. Naturally, trial solutions for B sites are expressed as

$$\beta_{m,n}(t) = \rho e^{i\phi} p(x,y,\tau,T) + \rho^2 [q_0(x,y,\tau,T) + e^{i\phi} q_1(x,y,\tau,T) + e^{2i\phi} q_2(x,y,\tau,T)]$$
$$+ \rho^3 [r_0(x,y,\tau,T) + e^{i\phi} r_1(x,y,\tau,T) + e^{2i\phi} r_2(x,y,\tau,T) + e^{3i\phi} r_3(x,y,\tau,T)] + \ldots$$

(8)

By inserting ansatzes (10) and (11) into those equations of motion on coherent-state amplitudes and equating coefficients of each harmonic frequency at each order of $\rho$, we can get two sets of equations.

According to the $O(\rho e^{i\phi})$ terms of Eqs. (4)–(5), one can get two equations relating $f$ and $p$, i.e.,

$$\mathbf{M} \begin{pmatrix} f \\ p \end{pmatrix} := \begin{pmatrix} \omega_0 - \omega - 2J_2 S \cos(k_y) & -4S(J_1 \gamma_k + iDm_k) \\ 4S(-J_1 \gamma_k + iDm_k) & \omega_0 - \omega - 2J_2 S \cos(k_x) \end{pmatrix} \begin{pmatrix} f \\ p \end{pmatrix}$$
$$= 0,$$

(9)

with

$$\gamma_k = \cos(k_x)\cos(k_y), \quad m_k = \sin(k_x)\sin(k_y) \tag{10}$$

Considering that we pay attention only to untrivial solution, thus Eq. (9) is in fact an common eigenvalue problem, with the eigenvalue $\omega$. Making use of the corresponding secular equation, it is not difficult to give the magnon dispersion relation with the following form

$$\omega = \omega_0 - J_2 S[\cos(2k_x) + \cos(2k_y)] \\ \pm S\sqrt{J_2^2[\cos(2k_x) - \cos(2k_y)]^2 + 16(J_1^2 \gamma_k^2 + D^2 m_k^2)}, \quad (11)$$

where the minus sign is corresponding to an acoustic branch, i.e., surface in $(k_1, k_2, \omega)$ space of lower frequencies, and the surface corresponding to the plus sign is referred to as the optical branch. When the present checkerboard ferromagnet does not possess the DM interaction, i.e., $D = 0$, the optical "up" and acoustic "down" frequency band meet at the corner of the Brillouin zone, as displayed in Fig. 3(a). Once the DM interaction is introduced, the spatial inversion symmetry of the two-dimensional checkerboard ferromagnetic lattice shall be destroyed so as to open a magnon band gap $\Delta\varepsilon = 8DS$ at the corner of the Brillouin zone, as shown in Fig. 3(b).

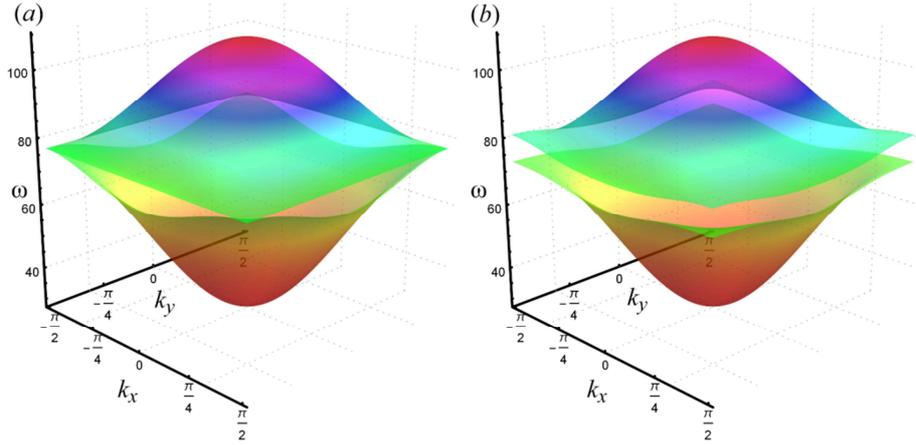

**Fig. 3.** The magnon dispersion surface for the two-dimensional checkerboard ferromagnet: (a) $D = 0$, (b) $D = 0.1$. The remaining parameters are set to $J_1 = 1$, $J_2 = 0.2$, $S = 10$, $H_{ext} = 10$, $A = 1$, $g = 1$, and $\mu_B = 1$.

It is not sufficient to get the frequency eigenvalue, we call for the form of the solution for $f$ and $p$ as well. Thus, we need to get the eigenvector of Eq. (9). Owing to $\det(\mathbf{M})=0$, here solutions are expressed as $(f,p)^T = f(1,C)^T$. By solving Eq. (9), one can get $C(k_1,k_2)$, which has the following form

$$C_{ac} = \frac{J_2 S[\cos(2k_x)-\cos(2k_y)]+\sqrt{J_2^2 S^2[\cos(2k_y)-\cos(2k_x)]^2+|\beta_k|^2}}{|\beta_k|}e^{i\theta},$$

$$C_{opt} = -\frac{[\sqrt{J_2^2 S^2[\cos(2k_y)-\cos(2k_x)]^2+|\beta_k|^2}-J_2 S[\cos(2k_x)-\cos(2k_y)]]}{|\beta_k|}e^{i\theta}$$

(12)

where $\beta_k \equiv |\beta_k|e^{i\theta} = 4S(J_1\gamma_k + iDm_k)$.

From the $O(\eta^2 e^{0i\varphi})$ and $O(\eta^2 e^{2i\varphi})$ terms, we can get

$$g_0 = q_0 = 0, \quad g_2 = q_2 = 0.$$
(13)

Let us consider the governing equations at $O(\eta^2 e^{i\varphi})$, which can be recast into

$$\mathbf{M}\begin{pmatrix} g_1 \\ q_1 \end{pmatrix} = \begin{pmatrix} if_\tau + id_1 f_y + ic_k p_x + id_k p_y \\ ip_\tau + id_2 p_x + ic_k^* f_x + id_k^* f_y \end{pmatrix}$$
(14)

with

$$c_k = 4S\left[J_1 \cos(k_y)\sin(k_x) - iD\cos(k_x)\sin(k_y)\right],$$

$$d_k = 4S\left[J_1 \cos(k_x)\sin(k_y) - iD\cos(k_y)\sin(k_x)\right],$$

(15)

where $\mathbf{M}$ is the matrix shown in Eq. (9).

Due to $\det(\mathbf{M})=0$, an equation such as Eq. (14), which can be rewritten as $\mathbf{M}(g_1,q_1)^T = \mathbf{F}$, either does not have solutions, or exist a whole family of solutions on $(g_1,q_1)^T$. According to the Fredholm alternative[31], the existence of solutions is in connection with $\mathbf{F}$. Solutions can exist only if $\mathbf{F}$ appears in the range of the matrix $\mathbf{M}$. It is not hard to find that $\mathbf{n} = (1/C^*, 1)^T$ satisfies $\mathbf{n} \cdot \mathbf{F} = 0$ in both the optical and the acoustic branches.

Combining $p = Cf$ with $\mathbf{n} \cdot \mathbf{F} = 0$, Eq. (14) yields

$$f_\tau + \frac{|C|^2 d_2 + c_k^* C^* + c_k C}{1 + |C|^2} f_x + \frac{d_1 + d_k C + C^* d_k^*}{1 + |C|^2} f_y = 0 \tag{16}$$

which means that there exists the travelling wave solutions for $f$ and $p$. Therefore, $f$ and $p$ can be written as

$$f(x, y, \tau, T) \equiv f(z, w, T), \quad p(x, y, \tau, T) \equiv p(z, w, T), \tag{17}$$

where $z = x - u\tau$ and $w = y - v\tau$. From Eq. (16), we can obtain the group velocity $\mathbf{v}_g = u\mathbf{e}_x + v\mathbf{e}_y$, where

$$u = \partial \omega / \partial k_x = \frac{|C|^2 d_2 + c_k^* C^* + c_k C}{1 + |C|^2}, \quad v = \partial \omega / \partial k_y = \frac{d_1 + d_k C + C^* d_k^*}{1 + |C|^2}. \tag{18}$$

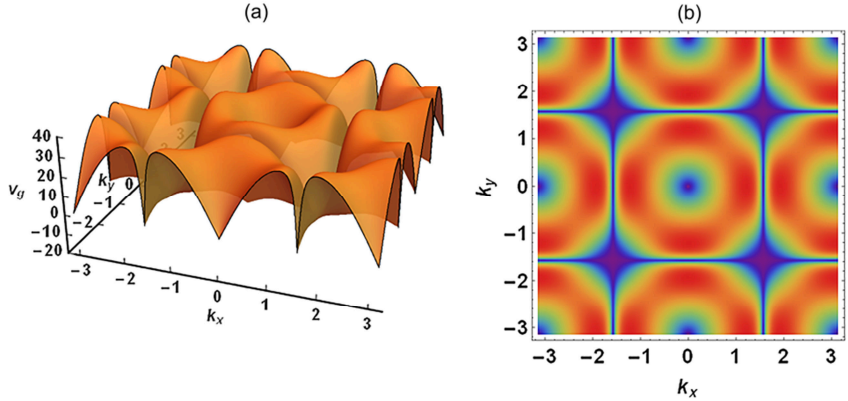

**Fig. 4.** (Color online) (a) The surface diagram for $v_g$; (b) The density diagram for $v_g$. The relevant parameters are $J_1 = 1$, $J_2 = 0.2$, $D = 0.1$, $S = 10$, $H_{ext} = 10$, $A = 1$, $g = 1$, and $\mu_B = 1$.

Obviously, the magnitude of the group velocity $\mathbf{v}_g$ has the following form

$$v_g = \sqrt{u^2 + v^2}. \tag{19}$$

Noting that the forms of $v_g^{ac}$ and $v_g^{opt}$ are same. Thus, one writes down

$v_g^{ac} = v_g^{opt} = v_g$ for any wave-vector $\mathbf{k}$. In Fig. 4, we display the magnitude of the group velocity $v_g$ as function of $(k_x, k_y)$. From Eq. (18), one can deduce that $u_{ac}$, $u_{opt}$, $v_{ac}$, and $v_{opt}$ are equal to zero at the center and corner of the Brillouin zone. Naturally, $v_g$ should be equal to zero at these points, as shown in Fig. 5(b).

In fact, the solution of Eq. (14) is degenerate, hence its one-parameter family of solutions may be written as

$$\begin{pmatrix} g_1 \\ q_1 \end{pmatrix} = \bar{g}_1 \begin{pmatrix} 1 \\ C \end{pmatrix} + \hat{g}_1 \begin{pmatrix} 1 \\ 0 \end{pmatrix}. \tag{20}$$

Here, $\hat{g}_1$ is determined by Eq. (14), and $\bar{g}_1$ is an arbitrary function. Since two generation equations for $\hat{g}_1$ from Eq. (14) are identical, $\hat{g}_1$ can be expressed as $\hat{g}_1 = i\hat{u} f_z + i\hat{v} f_w$, where

$$\hat{u} = \frac{s(c_k C - \frac{c_k^*}{C} - d_2)}{2\sqrt{J_2^2 S^2 [\cos(2k_x) - \cos(2k_y)]^2 + |\beta_k|^2}}, \hat{v} = \frac{s(d_1 + d_k C - \frac{d_k^*}{C})}{2\sqrt{J_2^2 S^2 [\cos(2k_x) - \cos(2k_y)]^2 + |\beta_k|^2}}. \tag{21}$$

Here, $s = \pm 1$, $d_1 = 4SJ_2 \sin(2k_y)$ and $d_2 = 4SJ_2 \sin(2k_x)$. It is pointed that $s = +1$ and $-1$ correspond to the optical and acoustic branches, respectively.

By considering terms at $O(\eta^3 e^{0i\varphi})$, one can obtain

$$h_0 = r_0 = 0. \tag{22}$$

By analyzing from terms of $O(\eta^3 e^{i\varphi})$, we can get the final equation, which is given by

$$\mathbf{M} \begin{pmatrix} h_1 \\ r_1 \end{pmatrix} = \begin{pmatrix} C_1 \\ C_2 \end{pmatrix}. \tag{23}$$

Here, the form of the matrix $\mathbf{M}$ has been given in Eq. (9), and the RHS components read

$$\begin{aligned}
C_1 = &\, i(g_{1\tau}+f_T)+4J_1 f|p|^2 - p|p|^2 (J_1\gamma_k+iDm_k)-\left[-2(A+J_2)+J_2\cos(k_y)\right]f^2 f^* \\
&-(J_1\gamma_k-iDm_k)f^2 p^* + 4SJ_2\cos(k_y)f_{yy} + 4iJ_2 S\sin(k_y)g_{1y} - 4SJ_1 m_k p_{xy} \\
&+2iSm_k Dp_{xx}+2iSm_k Dp_{yy}-4iS\gamma_k Dp_{xy}+2S\gamma_k J_1 p_{xx}+2S\gamma_k J_1 p_{yy} \\
&-4S\left[-iJ_1\cos\left(\frac{k_y}{2}\right)\sin\left(\frac{k_x}{2}\right)-D\cos\left(\frac{k_x}{2}\right)\sin\left(\frac{k_y}{2}\right)\right]q_{1x} \\
&-4S\left[-D\cos\left(\frac{k_y}{2}\right)\sin\left(\frac{k_x}{2}\right)-iJ_1\cos\left(\frac{k_x}{2}\right)\sin\left(\frac{k_y}{2}\right)\right]q_{1y},
\end{aligned}$$

$$\begin{aligned}
C_2 = &\, i(q_{1\tau}+p_T)+4J_1 p|f|^2 - (J_1\gamma_k-iDm_k)f|f|^2 - \left[-2(A+J_2)+J_2\cos(k_x)\right]p^* p^2 \\
&-(J_1\gamma_k+iDm_k)f^* p^2 + 4SJ_2\cos(k_x)p_{xx} + 4SiJ_2\sin(k_x)q_{1x} - 4Sm_k J_1 f_{xy} \\
&-2Sim_k Df_{xx}-2Sim_k Df_{yy}+4iS\gamma_k Df_{xy}+2S\gamma_k J_1 f_{xx}+2S\gamma_k J_1 f_{yy} \\
&-4S\left[-iJ_1\cos\left(\frac{k_y}{2}\right)\sin\left(\frac{k_x}{2}\right)+D\cos\left(\frac{k_x}{2}\right)\sin\left(\frac{k_y}{2}\right)\right]g_{1x} \\
&-4S\left[D\cos\left(\frac{k_y}{2}\right)\sin\left(\frac{k_x}{2}\right)-iJ_1\cos\left(\frac{k_x}{2}\right)\sin\left(\frac{k_y}{2}\right)\right]g_{1y}.
\end{aligned}$$

(24)

In order make sure that these equations exist solutions, we should use the consistency condition $\mathbf{n}\cdot\begin{pmatrix}C_1\\C_2\end{pmatrix}=0$. Taking $g_1$, $q_1$ as given via Eq. (20) with $\bar{g}_1=-\hat{g}_1/(1+|C|^2)$ and utilizing $p=Cf$, we can get

$$if_T + p_1 f_{zz} + p_2 f_{zw} + p_3 f_{ww} + q|f|^2 f = 0, \tag{25}$$

where

$$\begin{aligned}
p_1 &= \frac{1}{2}\frac{\partial^2\omega}{\partial k_x^2} \\
&= \frac{1}{1+|C|^2}\left[\frac{1}{2}\beta_k^* C^* + \frac{1}{2}\beta_k C + 2(\omega_0+\omega_x)|C|^2\right] - \frac{|C|^2 C^*}{(1+|C|^2)^2}c_k^*\hat{u} + c_k\frac{C}{(1+|C|^2)^2}\hat{u} + d_2\frac{|C|^2\hat{u}}{(1+|C|^2)^2},
\end{aligned}$$

$$\begin{aligned}
p_2 &= \frac{\partial^2\omega}{\partial k_x \partial k_y} \\
&= -\frac{1}{1+|C|^2}[4S(m_k J_1-i\gamma_k D)C^* + 4S(J_1 m_k+i\gamma_k D)C] \\
&\quad -c_k^*\frac{|C|^2 C^*}{(1+|C|^2)^2}\hat{v}+c_k\frac{C\hat{v}}{(1+|C|^2)^2}-\frac{C^*|C|^2}{(1+|C|^2)^2}d_k^*\hat{u}+\frac{C}{(1+|C|^2)^2}d_k\hat{u}-\frac{|C|^2}{(1+|C|^2)^2}d_1\hat{u}+d_2\frac{|C|^2\hat{v}}{(1+|C|^2)^2},
\end{aligned}$$

$$p_3 = \frac{1}{2}\frac{\partial^2 \omega}{\partial k_y^2}$$

$$= \frac{1}{1+|C|^2}[\frac{1}{2}\beta_k C + \frac{1}{2}\beta_k^* C^* + 2(\omega_0 + \omega_y)] - \frac{C^*|C|^2}{(1+|C|^2)^2}d_k^*\hat{v} + \frac{C}{(1+|C|^2)^2}d_k\hat{v} - \frac{|C|^2}{(1+|C|^2)^2}d_1\hat{v},$$

$$q = \frac{1}{1+|C|^2}\left\{8|C|^2 J_1 - \left[-2(A+J_2)+J_2\cos(2k_x)\right]\left(1+|C|^4\right) - 2(J_1\gamma_k - iDm_k)C^* - 2(J_1\gamma_k + iDm_k)|C|^2 C\right\}.$$

(26)

Obviously, Eq. (36) is the famous (2+1) nonlinear Schrödinger (NLS) equation.

### 4. Nonlinear localized modes

In our previous work[24], we have obtained birght and dark solition solutions to (2+1) NLS equation by using the Hirota bilinear method. In this section, we make use of these solition solutions to construct the analytical expressions on nonlinear localized modes. Here, we focus on two types of special wave vectors: (i) $\mathbf{k}=0$, (ii) $\mathbf{k}=(\pi/2,\pi/2)$.

#### A. The Brillouin zone center mode

At the Brillouin zone center (i.e., $\vec{k}=0$), the system may support an interesting nonlinear localized mode, namely, the Brillouin zone center modes. Since the magnon frequency spectrum of the checkerboard ferromagnet possesses two branches, the acoustic mode and optical mode at the Brillouin zone center shall be respectively condiered.

For the acoustic branch, one has $P_{1,ac}=P_{3,ac}=2J_1 S+2J_2 S$, $P_{2,ac}=0$, and $Q_{ac}=2J_1+J_2+2A$. Furthermore, we find that $\omega_{ac}=\omega_{\min}=\omega_0-2J_2 S-4J_1 S$, $u_{ac}=v_{ac}=0$, and $C_{ac}=1$. Thus, Eq. (25) can be written as the focusing 2D NLS equation, which exists a bright soliton solution. In this case, it is not hard to construct the analytical form of the Brillouin zone center acoustic mode, which is

$$\alpha_{m,n}(t) = \beta_{m,n}(t) \approx \varepsilon e^{-i\Omega_1 t} \frac{e^{a_1 \varepsilon m + a_2 \varepsilon n}}{1 + \frac{2J_1 + J_2 + 2A}{16S(J_1 + J_2)(a_1^2 + a_2^2)} e^{2\varepsilon(a_1 m + a_2 n)}}, \quad (27)$$

where $\Omega_1 = \omega_{min} - 2\varepsilon^2 S(J_1 + J_2)(a_1^2 + a_2^2)$. We note that Eq. (27) is a 2D intrinsic localized mode with the bright localized structure, whose eigenfrequency $\Omega_1$ lies below the bottom of the acoustic branch frequency band. Fig. 5(a) displays the square of the coherent-state amplitude of this 2D bright intrinsic localized mode, whose spatially localized structure can remain unchanged.

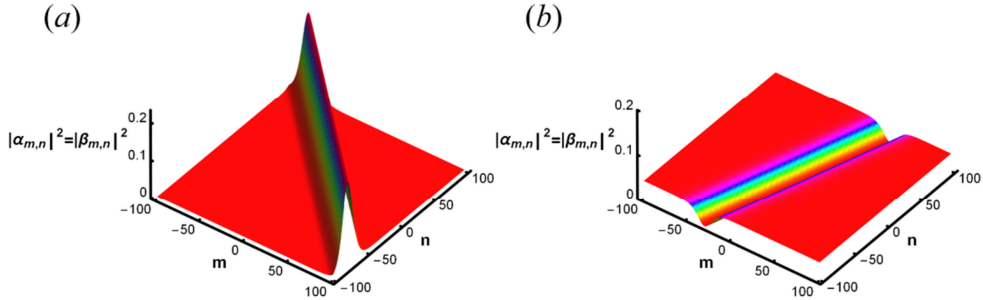

**Fig. 4.** (Color online) The Brillouin zone center modes in the checkerboard ferromagnet. (a) bright intrinsic localized mode: $\varepsilon = 0.1$, $a_1 = 1$, $a_2 = 1$; (b) dark intrinsic localized mode: $\varepsilon = 0.1$, $\eta = 2$, $b_2 = 0.6$.

Next, we turn our attention to the acoustic branch. For the acoustic branch, we have $P_{1,opt} = P_{3,opt} = -2S(J_1 - J_2) > 0$, $P_{2,opt} = 0$, and $Q_{opt} = 6J_1 + J_2 + 2A$. Then, Eq. (25) is the defocusing NLS equation, which supports a dark soliton solution. What is more, we note that $\omega_{opt} = \omega_{max} = \omega_0 + 3JS$, $u_{opt} = v_{opt} = 0$, and $C_{opt} = -1$. In this case, one can obtain the analytical expression of the optical the Brillouin zone center mode, namely,

$$\alpha_{m,n}(t) = -\beta_{m,n}(t) \approx \varepsilon \eta e^{-i\Omega_2 t} \tanh\left( \frac{\sqrt{S(J_1 - J_2)\left[(6J_1 + J_2 + 2A)\eta^2 - S(J_1 - J_2)b_2^2\right]}}{-2S(J_1 - J_2)} \varepsilon m + \frac{b_2 \varepsilon n}{2} \right), \quad (28)$$

where $\Omega_2 = \omega_{max} - \varepsilon^2 \eta^2 (6J_1 + J_2 + 2A)$. Obviously, Eq. (28) is a 2D intrinsic localized mode with a dark localized structure, as displayed in Fig. 4(b). In fact, this dark intrinsic localized mode is a resonant mode [32] since its vibration frequency lies within the magnon optical frequency band. Furthermore, it is in resonance with the optical magnon, which causes the finite lifetime of the 2D dark intrinsic localized mode in real ferromagnetic materials.

### B. The Brillouin zone corner mode

Now, let us consider nonlinear localized mode at the corner of the Brillouin zone. For the acoustic branch, we have $p_{1,ac} = p_{3,ac} = 2S(D - J_2)$, $p_{2,ac} = 0$, $q_{ac} = 4J_1 + 3J_2 + 2A - 2D$, $C_{ac} = -i$, $u_{ac} = v_{ac} = 0$, and $\omega_{ac} = \omega_0 + 2J_2 S - 4DS$. In this case, the 2D NLS equation (25) becomes

$$if_T + 2S(D - J_2)(f_{zz} + f_{ww}) + (4J_1 + 3J_2 + 2A - 2D)|f|^2 f = 0. \tag{29}$$

From the above equation, it can be seen that there is a competitive relationship between DM interaction and Heisenberg next-nearest neighbor exchange interaction, which causes that the Brillouin zone corner mode can have different local structures. In the case of $D < J_2$, Eq. (25) corresponds to a defocusing NLS equation, which exists a dark soliton solution. Thus, we can get the analytical expression of the Brillouin zone corner acoustic mode,

$$\alpha_{m,n}(t) = i\beta_{m,n}(t)$$
$$\approx \varepsilon \eta e^{i\frac{\pi}{2}(m+n)} e^{-i\Omega_3 t} \times$$
$$\tanh\left(\frac{\sqrt{S(J_2 - D)[S(D - J_2)b_2^2 + (4J_1 + 3J_2 + 2A - 2D)\eta^2]}}{2S(D - J_2)} \varepsilon m + \frac{b_2 \varepsilon n}{2}\right) \tag{30}$$

with $\Omega_3 = \omega_{ac} - \eta^2 (4J_1 + 3J_2 + 2A - 2D)\varepsilon^2$, which is the 2D dark intrinsic localized mode. In Fig. 5(a), we show the square of the coherent-state amplitude of this dark intrinsic localized mode, which has a finite lifetime due to due to resonance with the acoustic magnon.

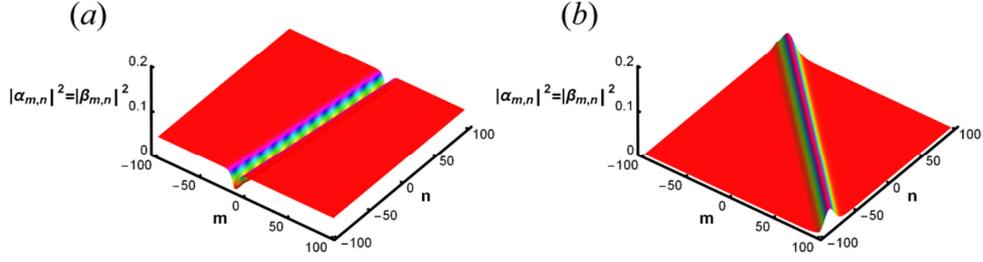

**Fig. 4.** (Color online) The Brillouin zone corner acoustic modes in the checkerboard ferromagnet. (a) $D = 0.1$; (b) $D = 0.5$.

When $D > J_2$, Eq. (25) is in fact a focusing NLS equation, which possess a bright soliton solution. Using the bright soliton solution in Rf. [24], it is easy to construct the analytical expression of the Brillouin zone corner acoustic mode,

$$\alpha_{m,n}(t) = i\beta_{m,n}(t) \approx \varepsilon e^{i\frac{\pi}{2}(m+n)} e^{-i\Omega_4 t} \frac{e^{a_1\varepsilon m + a_2\varepsilon n}}{1 + \frac{4J_1 + 3J_2 + 2A - 2D}{16S(-J_2 + D)(a_1^2 + a_2^2)} e^{2a_1\varepsilon m + 2a_2\varepsilon n}}. \quad (31)$$

where $\Omega_4 = \omega_{ac} - 2S(D - J_2)(a_1^2 + a_2^2)\varepsilon^2$ is within the magnon acoustic frequency band. It is Obvious that Eq. (31) is a 2D intrinsic localized resonant mode with the bright localized structure, as displayed in Fig. 4(b).

Based on the above analysis, it can be seen that the competition between the DM interaction and the Heisenberg next-nearest neighbor exchange interaction of the checkerboard ferromagnet has a great influence on the property of the Brillouin zone corner acoustic mode. If $D < J_2$, then the system supports the appearance of the dark intrinsic localized mode. When $D > J_2$, the bright intrinsic localized mode can emerge in the present checkerboard ferromagnet. In addition, we note that these Brillouin zone corner acoustic modes are resonant modes so that their lifetimes are finite.

Finally, we focus on the Brillouin zone corner optical mode. For the optical branch, we have $p_{1,opt} = p_{3,opt} = -2S(J_2 + D)$, $p_{2,opt} = 0$, $E_{opt} = 4J_1 + 3J_2 + 2A + 2D$, $C_{opt} = i$, $u_{opt} = v_{opt} = 0$, and $\omega_{opt} = \omega_0 + 2J_2 S + 4DS$. Thus, the 2D NLS equation (25) can be reduced to the following form

$$if_T - 2S(J_2 + D)(f_{zz} + f_{ww}) + (4J_1 + 3J_2 + 2A + 2D)|f|^2 f = 0, \qquad (32)$$

which is a defocusing NLS equation supporting the dark soliton solution. With our dark solition solution, one can get the analytical expression of the Brillouin zone corner optical mode, which is

$$\begin{aligned}\alpha_{m,n}(t) &= -i\beta_{m,n}(t) \\ &\approx \eta\varepsilon e^{i\frac{\pi}{2}(m+n)} e^{-i\Omega_5 t} \times \\ &\tanh\left(\frac{\sqrt{S(J_2+D)\left[-S(J_2+D)b_2^2 + (4J_1+3J_2+2A+2D)\eta^2\right]}}{-2S(J_2+D)}\varepsilon m + \frac{b_2\varepsilon n}{2}\right)\end{aligned} \qquad (33)$$

with $\Omega_5 = \omega_{opt} - \eta^2(4J_1 + 3J_2 + 2A + 2D)\varepsilon^2$. Eq. (33) is a dark type 2D intrinsic localized mode, as shown in Fig. 5. When the DM interaction strength $D$ satisfies the following relation

$$\frac{\eta^2(4J_1+3J_2+2A)\varepsilon^2 - 4J_2 S}{4S - 2\eta^2\varepsilon^2} < D < \frac{\eta^2(4J_1+3J_2+2A)\varepsilon^2}{4S - 2\eta^2\varepsilon^2}, \qquad (34)$$

the eigenfrequency $\Omega_5$ lies in the gap of two magnon frequency bands.

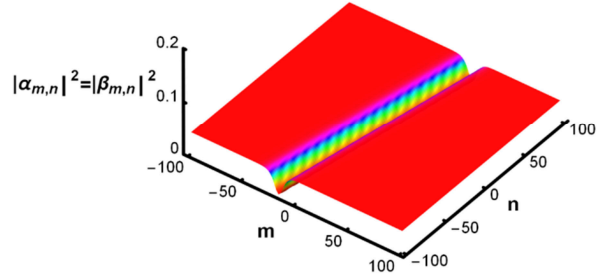

**Fig. 5.** (Color online) The Brillouin zone corner optical mode in the checkerboard ferromagnet.

## 5. Conclusions

In summary, the existence and property of the intrinsic localized mode in a two-dimensional checkerboard Heisenberg ferromagnet have been investigated in the semiclassical limit. With the help of an asymptotic method developed by Butt and Wattis, the governing lattice dynamical equations have been reduced to a (2+1)

–dimensional NLS equation. Using soliton solutions to the (2+1) –dimensional NLS equation, we have constructed different types of intrinsic localized mode solutions in the present checkerboard Heisenberg ferromagnet. Analytical forms for the Brillouin zone center mode and corner mode have been obtained. Their existence conditions and properties have been confirmed. Especially, our results have showed that the local structure of the Brillouin zone corner acoustic mode depends on the competition between the DM interaction and the Heisenberg next-nearest neighbor exchange interaction.

**Acknowledgments**

This work was supported by the National Natural Science Foundation of China under Grant Nos. 11875126, and 11964011, the Natural Science Fund Project of Hunan Province under Grant No.2020JJ4498.